\documentclass[prb,showpacs,showkeys,preprintnumbers,superscriptaddress,a4paper,twocolumn]{revtex4-1}
\usepackage{graphicx}
\usepackage{subcaption}
\usepackage{float}
\usepackage{esvect}
\usepackage{bbold}
\usepackage[T1]{fontenc}
\usepackage[utf8]{inputenc}
\usepackage{graphicx}
\usepackage{amsmath,amssymb}
\usepackage{amstext}													
\usepackage[colorlinks,citecolor=blue,filecolor=blue,linkcolor=blue,urlcolor=blue]{hyperref}

\newcommand{\uvec}[1]{\boldsymbol{\hat{\textbf{#1}}}}
\let\a=\alpha \let\b=\beta \let\g=\gamma

   \let\c=\chi

 \let\G=\Gamma

\def\beq{\begin{equation}}
\def\eeq{\end{equation}}
\def\bea{\begin{eqnarray}}
\def\eea{\end{eqnarray}}
\def\ba{\begin{array}}
\def\ea{\end{array}}

\begin{document}
\title{Quantum theory of spin waves for Helical ground states in Hollandite lattice.}
\author{Atanu Maity}\email{atanu.maity@iopb.res.in}\affiliation{Institute of Physics, Bhubaneswar-751005, Orissa, India}
\affiliation{Homi Bhabha National Institute, Mumbai - 400 094, Maharashtra, India}
\author{Saptarshi Mandal }\email{saptarshi@iopb.res.in}\affiliation{Institute of Physics, Bhubaneswar-751005, Orissa, India}
\affiliation{Homi Bhabha National Institute, Mumbai - 400 094, Maharashtra, India}

\begin{abstract}
 We perform  spin wave analysis of classical ground states of a model Hamiltonian proposed earlier(Phys. Rev. B {\bf 90}, 104420(2014)) for  
$\alpha-MnO_2$ compounds. It is known that the phase diagram of the Hollandite lattice (lattice of $\alpha-MnO_2$ compounds) consists of 
four different Helical phases(FH, A2H, C2H, CH phase)  in the  space of model parameters $J_1,~J_2,~J_3$. The spin 
wave dispersion shows presence of  gapless mode which interpolates between quadratic to linear depending on  phases and values of $J_i$'s.  
In most cases, the 2nd lowest mode shows the existence of a roton minima mainly from $X$ to $M$ and $M$ to $Z$ path. Few higher modes also 
show roton minima. Each helical phase has its characteristic traits which can be used to determine the phases itself.  The analytical 
expressions of eigenmodes at high symmetry points are obtained which can be utilized to extract the values of $J_i$'s. Density of states, 
specific heat and susceptibilities at low temperature has been studied within spin wave approximation. The specific heat 
shows  departure from  $T^{1.5(3)}$ dependence found in three dimensional unfrustrated ferromagnetic(anti-ferromagnetic) system which seems 
to be the signature of incommensurate helical phase. The parallel susceptibility is maximum for FH phase and minimum for CH phase at low temperature.  The perpendicular susceptibility is  found to be  independent of temperature at very low temperature. Our study can be  used to compare 
experimental results on  magnon spectrum, elastic neutron scattering, and finite temperature properties mentioned above for clean $\alpha-MnO_2$ system. 
\end{abstract}

\date{\today}

\maketitle
\section{Introduction}
Magnetic systems having geometrical frustration remains one of the key area of condensed matter research owing to their ability to produce diverse magnetic phases~\cite{satoru-1991,clark-1972,maegawa-1988,yelon-1973,kadowaki-1987,brener-1977,sano-1989} and other quantum many body effects like spin liquids~\cite{anderson-1973,fazekas-1974,tapp-2017,Piazza-2015},fractionalized excitations etc \cite{balents-2010,senthil-2000,saptarshi-2007}. These magnetic systems appear to posses spin magnetic moment  with $s=1/2$ to as large as $s=2$ in steps of $1/2$. The smaller the value of `$s$', the system behaves more quantum mechanically signifying that classical model is inadequate to describe such system. On the other hand, for large `$s$', the classical model is often sufficient to describe the system. Though in many cases finding the classical ground states is as difficult as finding its quantum counter part. Among various naturally existing materials that posses inherent magnetism, various Mn-based compounds are well known for long time \cite{sunil-1990,Ballou-1988,DeSavage-1965,kan-2013}. Among various Mn-based compounds, various Manganese-Oxides are known for their complex magnetic behaviour. They are also known for their diverse properties for example multiferroics and numerous applications in technology and agriculture. The lattice structures of these various $MnO_2$ are different and  some times the origin of frustration is due to the competing interactions along different bonds. In this respect, the material $\alpha-MnO_2$ which has elementary magnetic moment $s=3/2$, is note-worthy owing to their similarity with triangular lattice. This particular  lattice structure is known as Hollandite lattice  and this lattice can be constructed out of 3 dimensional triangular lattice by selectively removing bonds keeping the 3 dimensional character intact \cite{deguzman-1994,liu-2014,pistoia-1995}. One of the  main feature  of these  materials is that the two dimensional triangular lattices are folded in three dimension in a way that a channel like structure is formed and the diameter of these chanells are big anough to  accommodate large cations. This properties leads to many application  \cite{xiao-2010,khilari-2013,thackeray-1997,bruce-2012} as electrode materials in batteries, catalyst in oxygen reduction reactions etc. \\

The earilier studies on these particular $MnO_2$ was mainly confined to experiments which showed that these materials posses different magnetic phases depending on temperature as well as  impurity concentration.  For example, the experimental investigation in  Ref. \onlinecite{sato-1997} on $K_{1.5}MnO_8O_{16}$ resulted in three different phase transitions. First one (at temperature near 175K-2015K) is a charge ordering transition of $Mn^{4+}$ \& $Mn^{3+}$  coupled with the ordering of impurities ($K^+$ in this case) in the channels. The second one (near temperature 52K) is para-magnetic to ferromagnetic transition which persists upto some lower temperature near 20K where third transition occurs by disappearance of the ferromagnetic order. The works \cite{luo-jpc-114-2010,luo-j.app.phy-105-2009} on $\alpha-K_{x}Mn_8O_{16}$ shows divergences in both zero field cooled and field cooled susceptibilities and noticeable hysteresis for $0.10\leqslant x\leqslant 0.15$. However for $0.15\leqslant x\leqslant 0.17$ the sample behaves anti-ferromagnetically with Neel temperature around  25K. These works also show the presence of spin glass phases\cite{luo-jpc-114-2010,luo-j.app.phy-105-2009}.\\

   Recently,  theoretical model has been prescribed~\cite{andreanov-2013,saptarshi-2014}  in order to understand the magnetism in $\alpha-MnO_2$ system. In Ref. \onlinecite{andreanov-2013} an Ising model has been proposed on Hollandite lattice which included three different coupling constants $J_i,~i=1,2,3$ due to their geometrical/orientational differences. This simple model was successful to produced some key feature of the Hollandite lattice for example i) the results of frustration and classically degenerate ground state manifold and ii) the existence of four different collinear magnetic phases (namely A2H, CH, FH, C2H)  depending on the sign and magnitude of coupling constants. This simple model also  predicts the existence of a spin-glass phase in the presence of impurity defect in agreement with earlier experiments. In Ref. \onlinecite{saptarshi-2014}, the Ising model were replaced by a 3 dimensional Heisenberg model as the materials  does not show any particular preferential directions of spontaneous magnetization. This Heisenberg model on Hollandite lattice improved some of the results of Ising model for example it established the presence of four different helical ground states in the model parameter space. It may be mentioned that earlier study suggests that helical ground state may be the possible ground state for some materials \cite{sato-1997}. The calculated  ratio of  parallel and perpendicular susceptibility at zero temperature  agrees well with experiments. Given the important advancement made in the above to previous theoretical studies on possible magnetic phases in certain $\alpha-MnO_2$ materials, we look for low energy spin wave excitations and its consequences at low but finite  temperature specific heat and susceptibilities. Spin wave spectrum or excitation is an important physical process which can easily be experimentally determined in elastic neutron scattering experiments. Given the rich classical phase diagram of Hollandite lattice, we think our studies will be important in further determining the true magnetic nature of the $\alpha-MnO_2$ materials. Our plan of presentation is the following. In section \ref{sec1}, we describe the classical ground states found in the relevant parameter space as found in the earlier study and define the H-P transformation \cite{holstein-1940}  for evaluating spin-wave spectrum. In section \ref{sub-hk}, we discuss the magnon-spectrum in detail for each phases.The density of states for magnon and finite temperature specific heat are described  in section \ref{sub-nnn} and section \ref{sp-heat} respectively. The finite temperature susceptibility study is done in  section \ref{suscept} and we end up the article with discussion in section \ref{ana-res}. 
\begin{figure}
\includegraphics[width=0.75\linewidth]{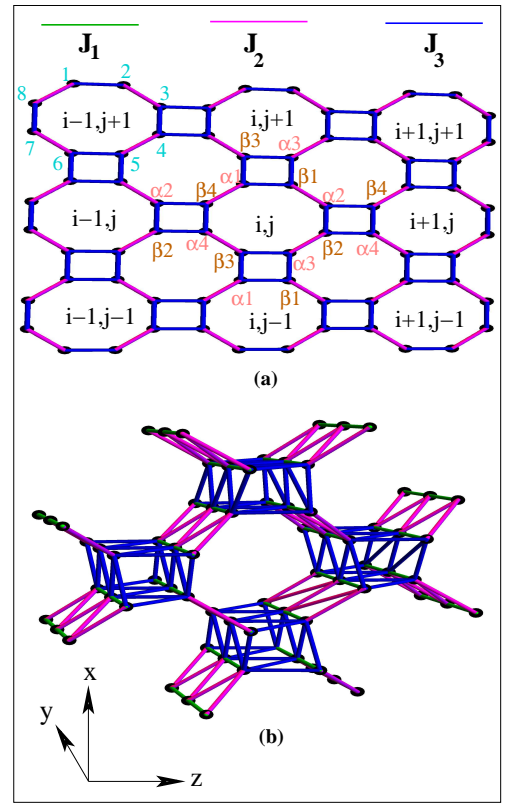}
\caption{Magnetic lattice structure of Hollandite lattice:  (a) projection in $x-z$ plane, (b) Three dimensional view of one octagonal channel surrounded by four square channel. Three different couplings, $J_1,J_2,J_3$ are shown using green, magenta and green lines respectively.}
\label{lattice}
\end{figure}
\section{Preliminaries: Hollandite lattice and Classical Ground States and spin-wave Hamiltonian}
\label{sec1}
\begin{figure}
\includegraphics[width=0.75\linewidth]{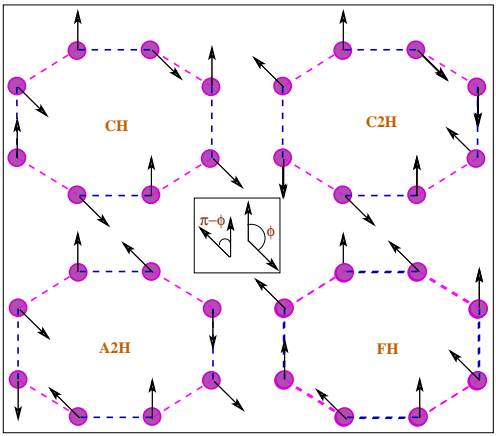}
\caption{Ground state configuration of unit cell for $|J_2|=|J_3|=J_1,\phi \approx 140^o$ in (a)CH, (b)C2H, (c)A2H \& (d)FH Phase.}
\label{four-phase}
\end{figure}

   In Fig\ref{lattice}, we schematically  showed  lattice structure of hollandite lattice. In actual material, each lattice site needs to be replaced by $Mn$-atoms having a tetrahedral structure with $O$-atoms. Fig. \ref{lattice}-a shows the projection of the lattice in $x-z$ plane and  Fig.~ \ref{lattice}-b shows a three dimensional view where three different colors have been used to describe three different kind of bonds utilized in the Heisenberg model on Hollandite lattice. A given unit cell  consists of 8-sites as indicated in Fig. \ref{lattice}. This unit cell is repeated in $x-z$ planes and as well as in $y$-direction. Each lattice site is connected by two $J_1$-type bonds in $y$-direction and we can call this $J_1$-chains. Two neighboring $J_1$ chains are connected by either $J_2$-bonds or $J_3$ bonds. Thus the Hollandite lattice can be visualized as  $J_1-J_2$ ladders interconnected by $J_3$ bonds or $J_1-J_3$ ladders interconnected by $J_2$ bonds. For easy description, we generically call the $J_1$ chains containing odd-sub-lattice (even-sub-lattice) as $\alpha_i$ ($\beta_i$) chains with  $i=1,2,3,4$. Thus we are having  four $\alpha$-chains \& four $\beta$-chains along y-direction. The relation of the subscripts '$i$' of the $\alpha/\beta$ chains to that of sub-lattice index $k$ is the following. For $\alpha$ chain $i=(2k-1)$ and for $\beta$ chain $i= 2k$. In an unit cell all the $\alpha$-sub-lattices are in same plane and all the $\beta$-sub-lattices are in other plane shifted by  constant displacement in `$y$' direction. The triangular ladder structure along the channel is the origin of geometrical frustration and non-trivial ordered phases in this system. The classical Heisenberg model on Hollandite lattice yields  various ground states depending on the signs and strength of $J_i$'s. In the classical  model $J_1$ has  always taken antiferromagnetic as the ferromagnetic choice renders the magnetic ground states as trivial as no frustration builds up in the system. For various values and magnitudes of $J_2$ and $J_3$, the magnetic ground states can be collinear or helical. There are four type of collinear or helical structures and they are called C-type (when $J_2$ and $J_3$ are both AFM), C2-type (when  $J_2$ is FM and $J_3$ are AFM), FM structure (when $J_2$ and $J_3$ are both FM)
and A2-type AFM (when $J_2$ is AFM and $J_3$ FM).  Various different kind of AFM phases differ from each other by mutual relative orientation among $\alpha$ (or $\beta$) chains as described below and shown in Fig.~\ref{four-phase}. To describe it we denote the ground state spin configuration of $\alpha_1$ and 
$\beta_1$ chains by $S_{\alpha_1}=\mathcal{S}$ and $S_{\beta_1}= \mathcal{T}$. Then various magnetic ground states can be explained as follows. \\

\begin{itemize}
\item{ {\bf{C-type}} Here the spin configuration of all  $\alpha$ chains are parallel to each other, i.e, $S_{\alpha_1}=S_{\alpha_2}=S_{\alpha_3}=S_{\alpha_4}=\mathcal{S} $.  Similarly for $\beta$-chains $S_{\beta_1}=S_{\beta_2}=S_{\beta_3}=S_{\beta_4}=\mathcal{T} $. The angle between spins in $\beta$ and $\alpha$  chains are $\phi$, $\phi=0,\pi$ for collinear phase.}
\item{ {\bf{A2-type}}  Here alternative $\alpha$-chains are anti-parallely aligned implying $S_{\alpha_1}=S_{\alpha_3}=\mathcal{S}=-S_{\alpha_2}=-S_{\alpha_4}$. Similar arrangments repeat for  $\beta$ chains.}
\item{ {\bf{C2H-type}} This one repeats the structures of  A2-type with the difference that while $\alpha$ chains repeats
the orientations of A2-type, the  $\beta$  chains are rotated by $\pi$ in $x-z$ plane relative to A2-type $\beta$ chains.}
\item{{\bf{FH-type}} This one repeats the structures of  C-type with the difference that while $\alpha$ chains repeats
the orientations of C-type, the  $\beta$  chains are rotated by $\pi$ in $x-z$ plane relative to C-type $\beta$ chains.}

\end{itemize}

\begin{figure}[htb]
\begin{center}
\includegraphics[scale=1]{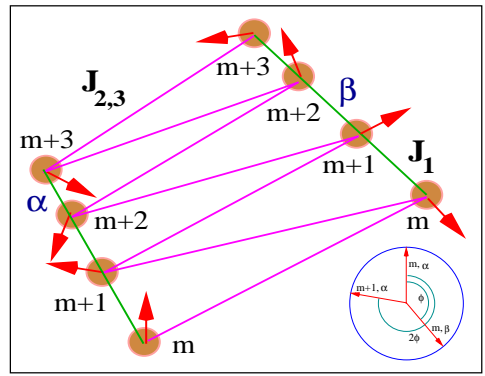}
\end{center}
\caption{The ground state configuration of an arbitrary $J_1-J_{2,3}$ Ladder. The red arrows are the spin orientations of this ladder for $J_{2,3}=3J_1 (\phi \approx 140^o)$}
\label{ladder}
\end{figure}

The above description signifies that one can write down the spin configuration of a given $\alpha$ chain (say for $\alpha_1$) and a given $\beta$ chain (say $\beta_1$) to describe the ground state spin-configuration according to the definition given above. In below we write down the ground state configuration  of this $\alpha-\beta$ chains as follows,  For Hollandite lattice this constitutes a $J_1-J_3$ ladder as shown in Fig.~\ref{ladder}. In CH phase, the ground state configuration can be written as
\begin{eqnarray}
\vec{S}_{m,\alpha}&=&s\left( \uvec{x}\cos{2m\phi}+\uvec{z}\sin{2m\phi} \right) \nonumber \\ 
\vec{S}_{m,\beta}&=& s\left(\uvec{x}\cos{(2m+1)\phi}+\uvec{z}\sin{(2m+1)\phi} \right) 
\label{alpha-beta-1}
\end{eqnarray}

Here the index `$m$' indicates the atomic sites and $\phi$ is the screw angle. Now we are in a position to perform Holstein-Primakoff transformations \cite{holstein-1940} for the classical ground states described above. To perform the H-P transformation for such non-colinear  magnetic ground state it is useful to have a relation between the local (denoted by $(x^{\prime}, y^{\prime}, z^{\prime})$) and global or laboratory co-ordinate system (denoted by $(x, y, z)$) where the quantization direction of a given spin defines the $x^{\prime}$-axis (which coincides with local spin alignment) of local co-ordinate system. The local and global co-ordinate system is connected via a rotation in $x-z$ plane as described below. 
 
\begin{eqnarray}
\begin{bmatrix}
{S^x}_{m,i} \\
{S^z}_{m,i}
\end{bmatrix}
=
\begin{bmatrix}
\cos{{\psi}_i} & -\sin{{\psi}_i} \\
\sin{{\psi}_i} & \cos{{\psi}_i}
\end{bmatrix}
\begin{bmatrix}
{S^{x'}}_{m,i} \\
{S^{z'}}_{m,i}
\end{bmatrix}
\label{alpha-beta-2}
\end{eqnarray} 
 
 Where, $i=\alpha ,\beta$. ${\psi}_\alpha=2m\phi$ and ${\psi}_\beta=(2m+1)\phi$.
 
The spin component $S^{\alpha^{\prime}}$ is defined as usual, 
\begin{eqnarray}
{S^{x'}}_{m,(\alpha/\beta)}&=&s-{a^\dagger}_{m,(\alpha/\beta)}  {a}_{m,(\alpha/\beta)} \nonumber \\
{S^{y'}}_{m,(\alpha/\beta)}&=&\sqrt{\frac{s}{2}}\left({a^\dagger}_{m,(\alpha/\beta)}+{a}_{m,(\alpha/\beta)}\right) \nonumber \\
{S^{z'}}_{m,(\alpha/\beta)}&=&i\sqrt{\frac{s}{2}}\left({a^\dagger}_{m,(\alpha/\beta)}-{a}_{m,(\alpha/\beta)}\right)
\label{spin-1}
\end{eqnarray}
Where ${a^\dagger}_{m,(\alpha/\beta)}$ $\&$  ${a}_{m,(\alpha/\beta)}$ represents creation $\&$ annihilation operator for magnon at $m^{th}$ site in $\alpha$ $\&$ $\beta$ chain respectively. The Heisenberg model on the Hollandite lattice can be written as
\begin{equation}
\label{hzero}
H = \sum_{k=1}^{3}\sum_{<i,j>}{J_k}^{ij}\vec{S_i}.\vec{S_j}
\end{equation}
Our Hamiltonian has two part, first it contains the interactions between nearest neighbor $\alpha$ and $\beta$ chains within a unit cell (intra-cell interaction) and secondly there are inter-cell interaction due to $J_1-J_3$ ladders which connect a unit cell with neighboring unit cells. After substituting the ground state ansatz for different phases as described before and implementing Eq.~\ref{alpha-beta-1}, Eq.~\ref{alpha-beta-2} and Eq.~\ref{spin-1}, we obtain the final Hamiltonian in momentum space as given below, 

\begin{equation}
H=\sum_{k(k_x,k_y,k_z)}\chi^\dagger_k M_k \chi_k + (\rm{constant\:~terms})
\label{mag-ham}
\end{equation}
where the basis,
\begin{eqnarray}
\chi_k &=& [\chi_{1,k}, \chi_{2,k}]^T \nonumber \\
\chi_{1k}&=&[a_{k,\a_1}..a_{k,\a_4}a_{k,\b_1}..a_{k,\b_4}] \\
\chi_{2k}&=&[{a^\dagger}_{-k,\a_1}..{a^\dagger}_{-k,\a_4}{a^\dagger}_{-k,\b_1}..{a^\dagger}_{-k,\b_4}]
\end{eqnarray}
and the coefficient matrix $M_k$ is a $(16\times 16)$ matrix. The detailed expressions of $M_k$ for different phases have been given in the appendix. Below we describe the magnon spectrum obtained from the above Hamiltonian. 
\section{Spin-wave spectrum}
\label{sub-hk}

The spectrum of the Hamiltonian in Eq.~\ref{mag-ham} is found in the usual way ~\cite{colpa-1978,xiao-arxiv-2009}. The spectrum is quite rich  and bear different characteristic signatures for different helical phases. In Fig.~\ref{dispersion}, we plot the dispersion along high symmetry points, $\Gamma (0,0,0)$, $X(\pi,0,0)$, $M(\pi,\pi,0)$, $Z(0,0,\pi)$, $R(\pi,0,\pi)$ \& $A(\pi,\pi,\pi)$. In all the phases we see the presence of accoustic and as well as optical modes. However, one needs to be cautious to use the phrase `accoustic mode' as though it is gapless and starts at the center of Briollouin zone (0,0,0), due to the helical angle $\phi$ which is not zero, the dispersion is not linear always. Such acoustic branches has been termes as quasi-acoustic branches in recent studies \cite{arakawa-arxiv-2017}). Also the appearance of roton minima in the 2nd lowest band in the magnon dispersion is one of the remarkable feature of the model. As the different phases has unique characteristic feature we prefer to describe the dispersion for different phases separately which will facilitate to get a comparative picture of all the phases.

\subsection{FH-phase}
\label{fh-dispersion}

  We begin our discussion with FH-phase which seems to be a good starting point to compare all other phases.  In this phase 
there is a gapless quadratic  mode at $\Gamma$ points which extends upto  $X$ point. The other prominent feature of this phase
is the existence of  8 non-degenerate bands (for the path $\Gamma$ to $X$ point) which are symmetrical in regard to the touching 
points of fourth and fifth bands. This unique feature is not repeated in other phases. From $X$ to $M$, the bands become doubly 
degenerate and all bands converge at $M$ points. In between $M$ to $Z$, the bands become non-degenerate but again converge at four almost 
equidistant eigenvalues. Form $Z$ to $R$, they are mostly four doubly degenerate bands which monotonously converge at a given eigenvalue at $A$ point.  
\subsection{A2H-phase}
\label{a2h-phase}
Now we turn to A2H phase. We remind that A2H phase differs from FH phase such that the alternative $\alpha$ chains are now alternatively antiferromagnetically arranged. Same happens for $\beta$ chains. This AFM alignment reflects in the fact that the gapless mode at $\Gamma$ points become more linear as clear from the 2nd right panel of the Fig.~\ref{dispersion}. This  feature continues for CH and C2H phase as well where there are some antiferromagnetic arrangements between different $\alpha$ (or $\beta$) chains. The other notable feature is that at $X$, $Z$ and $R$ points the 8 eigenvalues meet at three separate points only. The band width i.e the energy different from the highest to lowest also decreases compared to FH phase. 
\subsection{C2H-phase}
\label{c2h-phase}
The main characteristic feature of C2H phase is that the band width further decreases compared to A2H phase. However while in A2H phase  there are three
distinct eigenvalues at $Z$ and $R$ point, in C2H phase there are two low energy eigenvalue and two high energy eigenvalue
separated by a large gap appears. The acoustic mode at $M$ point represented by red plot also becomes 4 fold degenerate
while in A2H phase and FH phase it is two fold degenerate.
\subsection{CH-phase}
\label{ch-phase}
In CH-phase the band width is the minimum among all the four phases. However a unique characteristics is that the bands
are symmetric for the path $X~ \rightarrow ~M$ and $R~ \rightarrow~ A$. Also the eigenvalues are closely spaced than all other phases.
In the appendix we have given the Hamiltonian for different phases which can be used to get the spectrum for different phases.
From the material point of view this is very useful in the sense that one can use those formulas and play with various values of $J_i$
to match the experimental magnon spectrum. Further in Appendix B, we have given the exact expression for energy at specific symmetry point mainly at $\Gamma,~~X$ and $R$ point which can be used to compare with experimental data obtained from elastic neutron scattering.


\begin{figure*}
  \includegraphics[width=\textwidth,height=10cm]{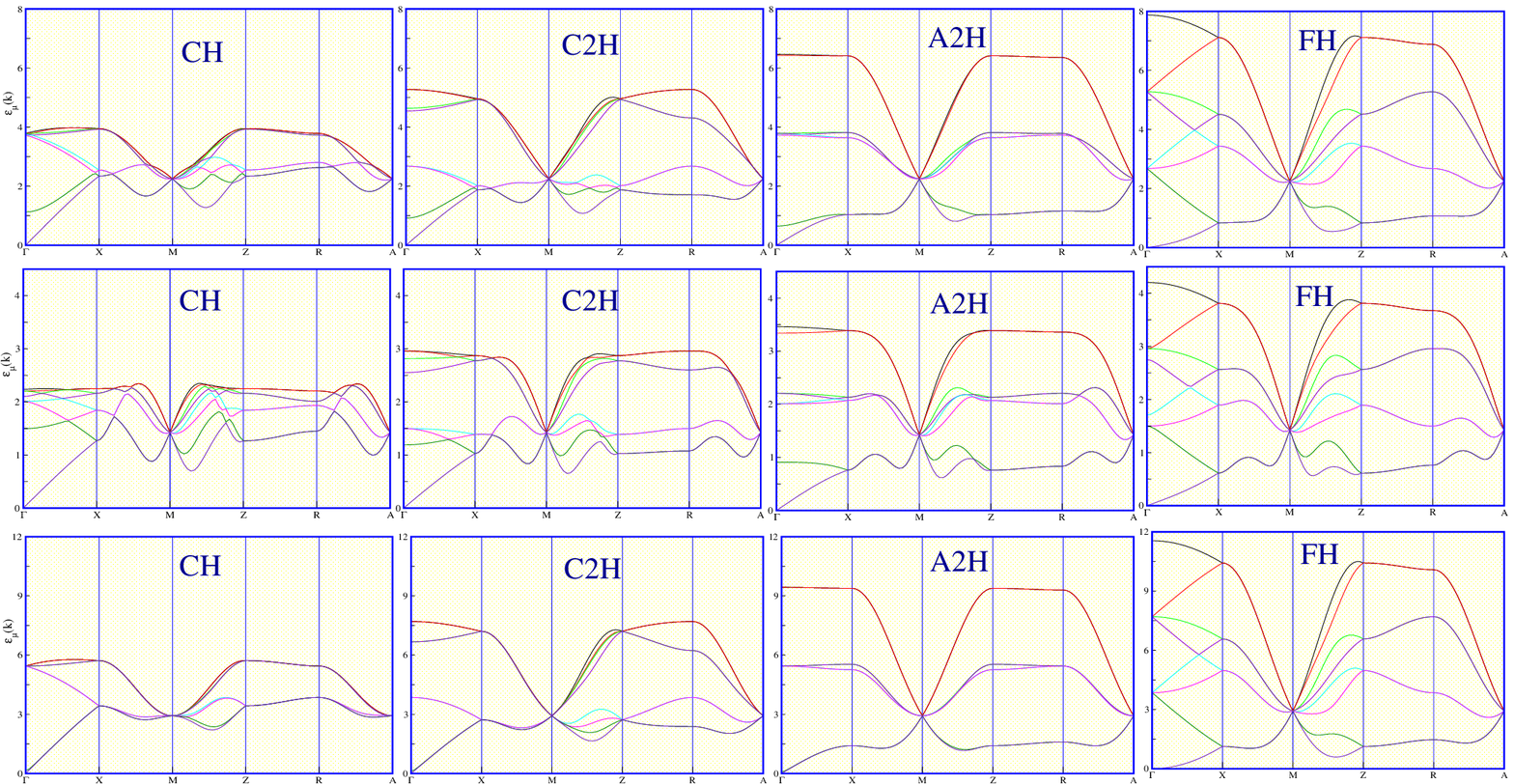}
  \caption{Spectrums for $|J_2/J_1|=|J_3/J_1|=1$ ($1^{st}$ row), $|J_2/J_1|=0.7, |J_3/J_1|=0.6$ ($2^{nd}$ row) \& $|J_2/J_1|=|J_3/J_1|=1.3$ ($3^{rd}$ row) where $\Gamma(0,0,0)$, $X(\pi,0,0)$, $M(\pi,\pi,0)$, $Z(0,0,\pi)$, $R(\pi,0,\pi)$, 
$A(\pi,\pi,\pi)$ are the symmetry points.}
\label{dispersion}
\end{figure*}

\begin{figure*}
  \includegraphics[width=\textwidth,height=4cm]{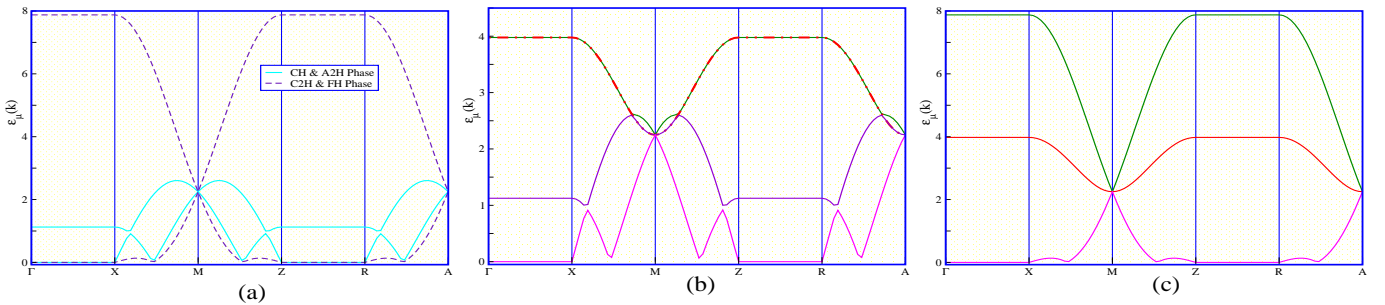}
  \caption{Spectrum for (a) $|J_2|=J_1,|J_3|=0$, (b) $|J_3|=J_1,|J_2|=0$ in CH \& C2H-Phase, (c)  $|J_3|=J_1,|J_2|=0$ in FH \& A2H-Phase; where $\Gamma(0,0,0)$, $X(\pi,0,0)$, $M(\pi,\pi,0)$, $Z(0,0,\pi)$, $R(\pi,0,\pi)$, 
$A(\pi,\pi,\pi)$ are the symmetry points.}
\label{chains}
\end{figure*}
 Also note that the excitation energies are 8-fold degenerate at $M(\pi,\pi,0)$ point. At this point the Hamiltonian contains only those terms which are related to individual chains and other inter chain coupling terms are not present, i.e., spin deviations are distributed along `$y$' direction only and no modes are available along $x$ \& $z$ direction at this point. The ground state configurations are identical for all individuals chains. It may be the reason behind the same excitation energies for fluctuations around $k_y=2\phi$ in all modes and all phases at this point. Excitation energy at this point is given by $\epsilon_\mu(k=(0,\pi,0))=S\sqrt{(E_{GS}+1)(E_{GS}+cos2\phi)}$ (Where $E_{GS}$ is the ground state energy per  site as obtained in the previous study using classical Heisenberg model~\cite{saptarshi-2014}) and we have checked that this is same whenever $k_y=\pi$ irrespective of  $k_x$ \& $k_z$ (i.e. $\epsilon_\mu(k_x,\pi,k_z)=\epsilon_\mu(0,\pi,0)$ ) as we can see at $A(\pi,\pi,\pi)$ points in the dispersion curves of Fig.~\ref{dispersion} and Fig. \ref{chains}.  \\
Before we  close the discussion we would like to mention that appearance of roton minima in the magnon dispersion is ubiquitous for this Hollandite lattice.
They mainly appear along the path of $X$ to $M$, $M$ to $Z$ and $R$ to $A$ points.  The datials nature of the roton minima and the band branches for which it appears depends on different phases. However given the analytical expressions of the Hamiltonian, it can easily be located for a given phase by varying the model parameters.


\section{DENSITY OF MAGNON MODES}
\label{sub-nnn}
\begin{figure}
\includegraphics[width=1.0\linewidth]{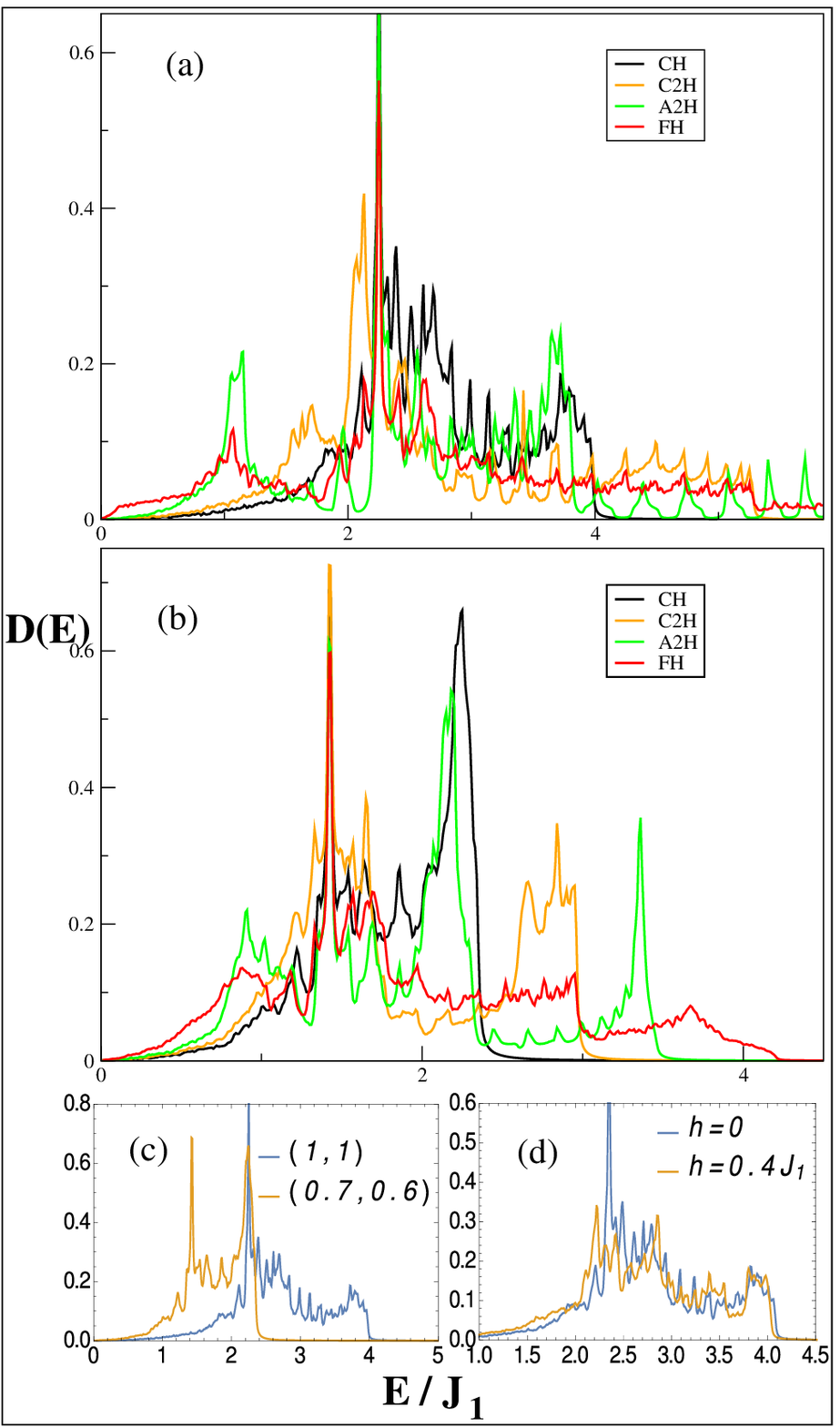}
\caption{Density of Magnon Modes in (a) all phases for $|J_2|=|J_3|=J_1$, (b) all phases for $(J_2/J_1,J_3/J_1)=(0.7,0.6)$, (c) CH phase for different $(J_2/J_1,J_3/J_1)$, (d) CH phase for different field for $|J_2|=|J_3|=J_1$.}
\label{dens-state}
\end{figure}

Now we briefly describe the profile of the density of states of different phases. The density of states is defined as the $\rm{tr(ImG(E))}$ where the Green function $G(E)$ is obtained from  $G_{\alpha_i\alpha_j}(k,\sigma)$ in the following way, 
\begin{eqnarray}
G_{\alpha_i\alpha_j}(k,\sigma) &=&-\left<T_\sigma a_{k,\alpha_i}(\sigma){a^\dagger}_{k,\alpha_j}(0)\right>_{\alpha\in \{\alpha,\beta\}} \\
G_{\alpha_i\alpha_j}(k,i\omega_n)& = &\int_{0}^{\beta} e^{i\omega_n\sigma}G_{\alpha_i\alpha_j}(k,\sigma)d\sigma
\end{eqnarray}
where the Matsubara Bosonic frequency, $\omega_n=\frac{2\pi{n}}{\beta}$. the density of states is obtained as,
\begin{equation}
D(E) =-\frac{1}{N\pi}\sum^{i=1,4}_{k}\rm{Im} \left( G_{\alpha_i\alpha_i}(k,E) + G_{\beta_i\beta_i}(k,E) \right)     
\end{equation} 
 Fig.~\ref{dens-state} represents the  plots for density of magnon modes in arbitrary unit for few parameter values. The plots shows that for low energy the density of states is vanishingly small due to the presence of low energy gapless mode which is constraint to have one dimensional character. The first peaks near $E=1$ is due to appearance of first gaped mode as evident from Fig.~\ref{dispersion}. The biggest peaks in density of states appears near $E=2.25$ which points to the excitations of $M$ point energy. A decrease of gap can be observed when we apply magnetic field for a particular choice of point in the phase diagram, see Fig.~\ref{dens-state}(d).  We also notice that for FH phase we have finite density of states at comparatively high energy and it gradually decreases for  A2H, C2H and CH phases. The density of states profile in conjunction with the dispersion spectrum described in earlier section will be useful for
determination of a particular phase.


\section{Low temperature specific heat}
\label{sp-heat}
\begin{figure}[h!]
\includegraphics[width=1.0\linewidth]{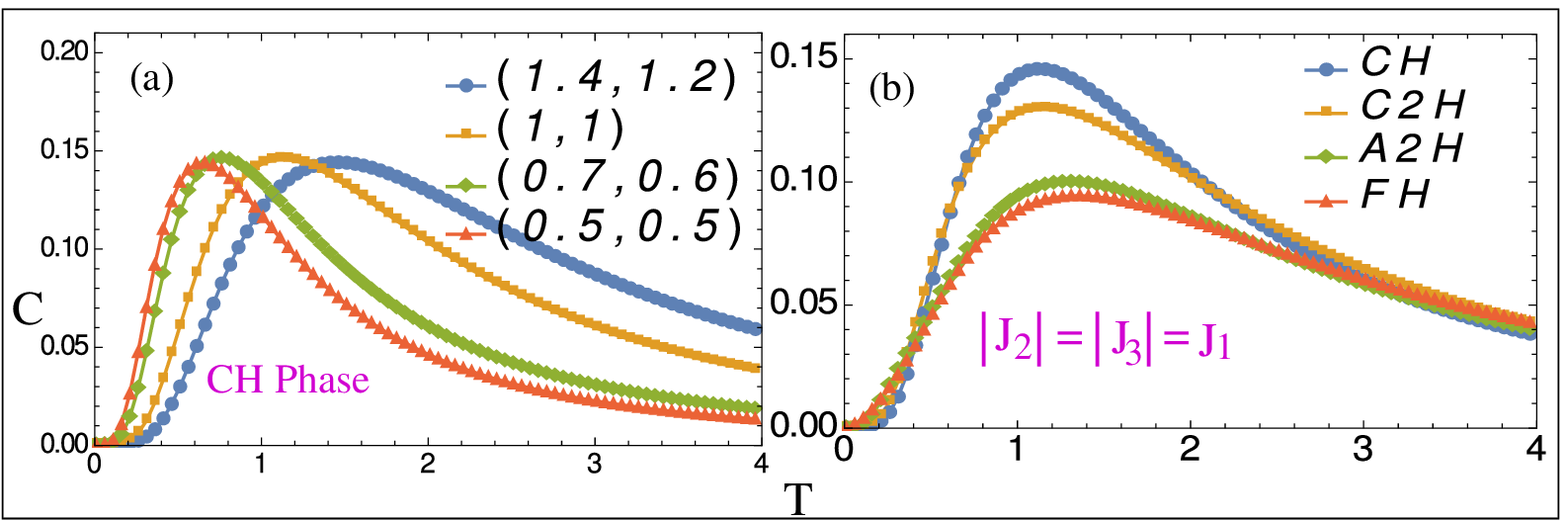}
\caption{ Magnon contribution to heat capacity for (a) different coupling strengths, $(J_2/J_1,J_3/J_1)$ in C-H Phase and (b) different phases for $J_2=J_3=J_1$.}
\label{sp-heat-1}
\end{figure}

\begin{figure}[h!]
\includegraphics[width=1.0\linewidth]{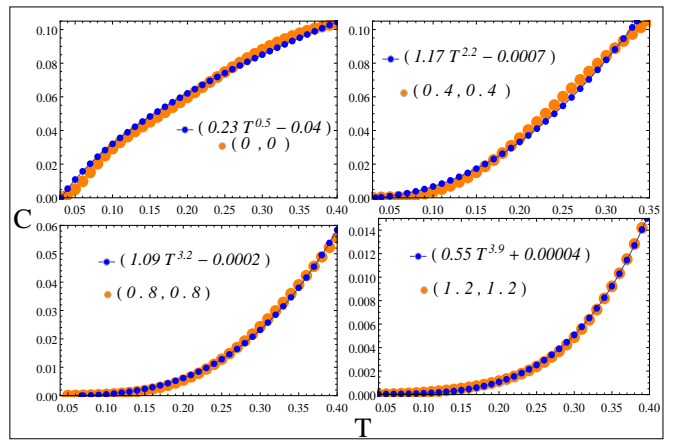}
\captionof{figure}{Different exponent dependence of Magnon contribution to heat capacity on temperature for different coupling strengths, $(J_2/J_1,J_3/J_1)$ in C-H Phase at extremely low temperature.}
\label{sp-heat-2}
\end{figure}

\begin{figure}[h!]
\includegraphics[width=1\linewidth]{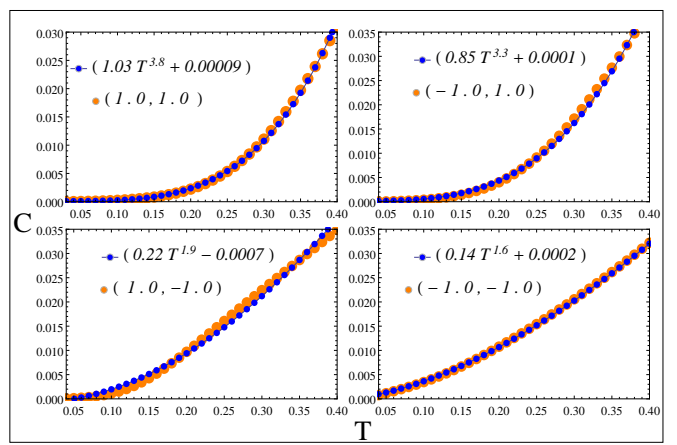}
\captionof{figure}{Different exponent dependence of Magnon contribution to heat capacity on temperature for different Phases at extremely low temperature where $|J_2|=|J_3|=J_1$.}
\label{sp-heat-3}
\end{figure}

The total energy($E(T)$) at a given temperature $T$ can be written as $E=\sum_{k,j} \epsilon_{k,j} \langle n_{k,j}(T) \rangle$. where $  \langle n_{k,j}(T) \rangle$ is the B-E distribution function and $ \epsilon_{k,j}$ is the eneragy for a given mode $j$ and wave vector $k$. This allowas us to write the expression of specific heat as given below,
\begin{eqnarray}
\label{sp-heat-eq}
C(T)&=&\frac{1}{k_BT^2} \sum_{k(k_x,k_y,k_z),j} \frac{{\epsilon_{k,j}}^2 e^{\beta \epsilon_{k,j}}}{(e^{\beta \epsilon_{k,j}}-1)^2} \\
    &=& \frac{1}{T^2} \int_{-\infty}^{\infty} D(E) dE \frac{E^2 e^{ E/T}}{(e^{E/T}-1)^2}
\end{eqnarray}
In the  second line of the above equation, we have converted the momentum summation to the integral over energy which is scalled by $J_1$ i.e,  $T\Rightarrow \frac{k_BT}{J_1}$. The above relation for specific heat due to excitations of bososnic quasi-particles can not be applied directly for  the present case  of magnon excitation where there is a constraint over magnon excitation number which is 2S for spin S. Therefore we obtained  specific heat from Eq. \ref{sp-heat-eq} using semi-classical approximation where we have assumed that at very low temperature, thermal excitations are much lower due to presence of effective gap.  Fig.~\ref{sp-heat-1}, Fig.~\ref{sp-heat-2} and Fig.~\ref{sp-heat-3} represent the low temperature dependence of  specific heat due to magnon contribution. The noticeable characteristic of these curves is that they exhibit anomalous peaks  known as Schottky anomaly\cite{gopal-1966,yarmohammadi-2013}. We expect this  peaks for a system having gap and it appears when thermal energy reaches the gap. Fig.~\ref{sp-heat-1} shows that the Schottky peak  shifts towards higher temterature which is due the increase of gap with the increase of coupling strengths. The Fig.~\ref{sp-heat-1}(b) also contains the specific heat plots for four similar points ($(|J_2/J_1|,|J_3/J_1|)=(1,1)$) in the phase diagram~\cite{saptarshi-2014} which shows that for a given set of model parameters, the peaks appear at same temperature though the peak height gradually decreases from  FH $\rightarrow$ A2H $\rightarrow$ C2H $\rightarrow$ CH phase.  Though  at higher temperature the crossover of specific heat  happens among various phases due to the detail changes of profile of density of states.  It may be noticed that the peak height is decreasing in phases other than CH phase due to reduction in peak height in density of state curves which points to the fact that in CH phase the energy bands are closely spaced than other phases as  evident from Fig.~\ref{dispersion}. 
\\
 It is also instructive  to compare the temperature dependence of heat capacity at very low temperature for hollandite system  with other known magnetic system. For example,  at very low temperature an unfrustrated collinear ferromagnet and antiferromagnet, the magnon contribution to specific heat depends on temperature as $T^{3/2}$ \& $T^3$  respectively~\cite{Nolthing_Ramakanth_qtheory_mag}.  Fig.~\ref{sp-heat-2} shows the low temperature dependence of specific heat in CH phase for different $(J_2/J_1, J_3/J_1)$ in phase diagram. The magnon contribution to specific heat depends approximately on temperature as $T^{3.8}$ and the temperature exponents decreases (increases) when we decrease (increase) coupling strengths. Fig.~\ref{sp-heat-3}, shows very low temperature dependence of specific heat in other phases. The temperature exponent is decreasing as we go from $CH\rightarrow C2H\rightarrow A2H\rightarrow FH$, i.e, specific heat is increasing. This is because at very low temperature lowest energy mode gives most of the contributions. This lowest energy modes are more available when we go from $CH\rightarrow C2H\rightarrow A2H\rightarrow FH$ [see dispersion curves and density of state curves discussed in previous sections]. In the next we analyze our results for low temperature susceptibilities for magnetic field applied parallel and perpendicular to the plane of polarization.



\section{Susceptibilities studies}
\label{suscept}
In this section we will study the effect of external magnetic field on the helical ground states. First we consider the case where the magnetic field is applied in the plane of polarization of the ground state spin configuration and in second case the magnetic field is applied perpendicular to the plane of polarization. For our case, the plane of polarization of ground state spin configuration is  $x-z$ plane and without loss of generality we can take  the magnetic field to be along $x$-axis. Below we provide the details of susceptibilities for parallel magnetic field.

\label{num-res}
\subsection{Magnetic field parallel to the plane of spin polarization}
The Hamiltonian in the presence of magnetic field applied along $x$-axis is given below,
\begin{eqnarray}
\label{hpara}
H_{h\parallel}= H -  \sum S_{i,x} h.
\end{eqnarray}
In the above $H$ represents the Hamiltonian given in Eq.~\ref{hzero} and magnetic field is assumed to be $\vec{h}= h \vec{e}_x$. The detailed steps of arriving at the expression of susceptibilities is given in Appendix-C. Here we give the outline. In the first step, we apply  the definition of  spin operators as given by Eq.~\ref{spin-1} to the Hamiltonian in Eq.~\ref{hpara}. After the above substitution, 2nd term in Eq.~\ref{hpara} yields both quadratic and linear terms. First the quadratic terms are taken into account with $H$ and diagonalized. Then the linear terms are re-expressed in the diagonalized basis. After a translation of the diagonalized basis by a constant terms, the liner terms can be absorbed to obtain the total energy. The diagonalized basis has the following form (with a displacement in $k$ as $k\rightarrow k-\phi$ to maintain the same positive and negative eigenvalues):

\begin{eqnarray}
\begin{bmatrix}
a_{k-\phi,\alpha} \\ a_{k-\phi,\beta} \\ a_{k+\phi,\alpha} \\ a_{k+\phi,\beta} \\ {a^\dagger}_{-(k-\phi),\alpha} \\ {a^\dagger}_{-(k-\phi),\beta} \\ {a^\dagger}_{-(k+\phi),\alpha} \\ {a^\dagger}_{-(k+\phi),\beta}
\end{bmatrix}
&=&
\begin{bmatrix}
T_{11} & T_{12} & . &. &. &T_{18} \\
 T_{21} \\
 .\\
 .\\
 .\\
 .\\
 .\\
 T_{81} & T_{82} & . &. &. & T_{88} 
\end{bmatrix}
\begin{bmatrix}
b_{k,1} \\ b_{k,2} \\ b_{k,3} \\ b_{k,4} \\ {b^\dagger}_{-k,1} \\ {b^\dagger}_{-k,2} \\ {b^\dagger}_{-k,3} \\ {b^\dagger}_{-k,4}
\end{bmatrix}
.
\end{eqnarray}

The induced magnetization, $M_{\|}$ can be written as, $M_{\|}=\left<S^x\right>=M_1+M_2$ where $M_1$ contains the quadratic terms and $M_2$ contains the linear terms. After a straightforward evaluation we obtain the following expressions for $M_1$ and $M_2$ as follows.
\begin{eqnarray}
M_1&=& -\frac{1}{2}\sum_{k,j}<n_{k,j}>C_j \\
M_2&=& \sqrt{\frac{s}{2}} \sum_{j=1,8}\frac{1}{\epsilon_{\phi,j}}p_j{p_j}^\star\delta_{k,\phi} 
\end{eqnarray}
here, $C_j=t_{1,j}{t^\star}_{j,3}+e^{-i\phi}t_{2,j}{t^\star}_{j,4}+h.c.$  and $t$-elements form the unitary transformation matrix that transforms $'a'$-basis to diagonalized $'b'$-basis $\&$ the bosonic distribution function, $n_{k,j}=\frac{1}{e^{\beta  \epsilon_{k,j}}-1}$. 

\begin{figure}
\includegraphics[width=0.75\linewidth]{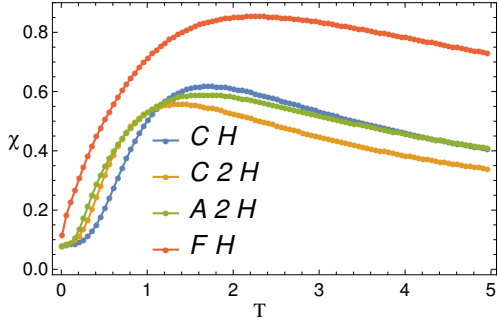}
\caption{ Magnetic Susceptibility for $|J_2|=|J_3|=J_1$ and $h=0.2J_1$ in all Phases.}
\label{sus1}
\end{figure}
\begin{figure}
\includegraphics[width=1.0\linewidth]{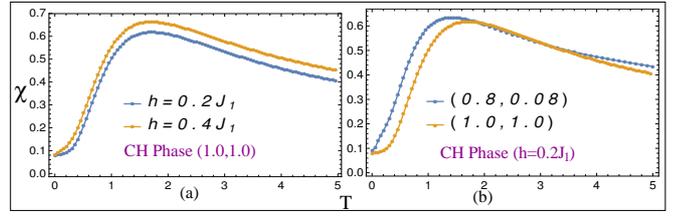}
\caption{Magnetic Susceptibility for (a) different field and (b) different coupling strengths, $(J_2/J_1,J_3/J_1)$ in CH phase.}
\label{sus-2}
\end{figure}

The Fig.~\ref{sus1} represents the magnetic susceptibility plots for $J_1=J_2=J_3$, indicating a low temperature magnetic ordering which persists upto some temperature and finally vanishes after further increase in temperature.  This results are also qualitatively supported by experimental results \cite{sato-1997,luo-jpc-114-2010,luo-j.app.phy-105-2009}. The Fig.~\ref{sus-2} shows the magnetic susceptibility for different field (a) and for different coupling strengths in CH phase (b). As expected we notice that for FH phase the magnetic susceptibility is maximum and gradually decreases as we go to A2H, C2H and CH phase. 

\subsection{Magnetic field perpendicular to plane of spin polarization}

Now we describe the effect of magnetic field having direction perpendicular to the plane of polarization, in our case this implies that the magnetic field
is applied in $y$-direction. The 2nd term in the Hamiltonian in Eq.~\ref{hpara} changes to  ($H_\bot =-h_\bot \sum_{m,j}{S^y}_{m,\a_j}$). In Fourier space this can be written as,
\begin{eqnarray}
{H_{\bot}}^\prime &=& \g \sum_{i,j}\left({a^\dagger}_{0,{\a}_j}+a_{0,{\a}_j}\right); \a , \b \in \a_j
\end{eqnarray}
Therefore total Hamiltonian in diagonalized basis can be written as,
\begin{multline}
H_\bot = \sum_{k,{\a}_j}\epsilon_{k,{\a}_j}{b^\dagger}_{k,{\a}_j}b_{k,{\a}_j}\\ +\g \sum_{i,j}\left( {t_{j,i}}^\star {b^\dagger}_{0,{\a}_j}+t_{i,j} b_{0,{\a}_j}\right)
\end{multline}
Where, $\g=-h_\bot \sqrt{\frac{s}{2}}$. Here we have performed a linear transformation( $b_{0,{\a}_j}\rightarrow b_{0,{\a}_j}+r_j\delta_{k,0}$) to recast the Hamiltonian into a quadratic forms in addition to a constant terms. For the proper choice of $r_j\left(=-\g \sum_i {t_{j,i}}^\star /\epsilon_{0,{\a}_j} \right)$, the Hamiltonian can be written as,
\begin{eqnarray}
H_\bot&=&\sum_{k,{\a}_j}\epsilon_{k,{\a}_j}{b^\dagger}_{k,{\a}_j}b_{k,{\a}_j}-\sum_{i,j}\frac{{\g}^2}{\epsilon_{0,{\a}_j}}|t_{j,i}|^2\delta_{k,0}
\end{eqnarray} 

Therefore the temperature dependence of magnetization can be written as, $M_{\bot}=<S^y>=\frac{1}{Z}tr\left(S^ye^{-\b H_\bot} \right)$ where, $ Z=tr\left(e^{-\b H_\bot}\right)$, is the partition function. The straight forward algebra yields the following expression for Magnetization and susceptibility:
\begin{eqnarray}
M_{\bot}&=&\sum_{i,j} \frac{{h_\bot} s{|t_{ij}|}^2}{\epsilon_{0,j}},~~{\c}_{\bot}=\sum_{i,j} \frac{ s{|t_{ij}|}^2}{\epsilon_{0,j}}
\end{eqnarray}
This expression shows that the perpendicular susceptibility is independent of temperature at low temperature, i.e, thermal energy do not effect the spin deviations along perpendicular direction at low temperature.
 
\section{Discussion}
\label{ana-res}

  To summarize we have found many new and insightful outcome of our spin wave analysis of clasical ground states proposed in the earlier study~\cite{saptarshi-2014}. The spin-wave spectrum shows  many unusual characteristics for example the existence of gapless modes which
can be either linear or quadratic depending on phases. This seems to be characteristics of Helical phases. The 2nd lowest modes shows
existence of roton minima for a large region of parameter spaces. Few of the higher modes also shows the roton minima. Each classical phases
has its own characteristics which can be useful in determining a particular phases. The analytical expressions of eigenvalues at high 
symmetry point can be used to determine the $J_i$'s. The finite temperature property for example specific heat and susceptibility shows
behavior similar to one dimensional system but the exact dependency on temperature differ from usual AFM/FM system. Our extensive study
can be useful  to compare with future experimental results. \\

{\bf{Acknowledgment}}:  We sincerely  thank  Mohsen Yarmohammadi for very useful discussion.

\section{APPENDIX-A}
\label{sub-hk}

The Magnon Hamiltonian has following form,
\begin{eqnarray}
M_k &=& s
\begin{bmatrix}
A(k) & B(k) \\
B(k) & A(k)
\end{bmatrix}
\end{eqnarray}
In the above we have droped the constant terms. The $A(k)$ and $B(k)$ are $8 \times 8$ matrix whose expression is given below
For magnetic field applied parallel to the plane of polarization, The full coefficient matrix (with $k_1=(k_x,k_y-\phi,k_z)$, and $k_2=(k_x,k_y+\phi,k_z)$ is the following,

\begin{eqnarray}
M_{k,h_{\parallel}} &=& s
\begin{bmatrix}
A(k_1) & F & B(k_1) & 0 \\
F^* & A(k_2) & 0 & B(k_2)\\
B(k_1) & 0 & A(k_1) & F \\
0 & B(k_2) & F^* & A(k_2)
\end{bmatrix}
\end{eqnarray}

where the basis,
\begin{eqnarray}
\chi_k &=& [\chi_{1,k_1},\chi_{1,k_2} \chi_{2,k_1},\chi_{2,k_2}]^T \nonumber \\
\chi_{1k_i}&=&[a_{k_i,\a_1}..a_{k_i,\a_4}a_{k_i,\b_1}..a_{k_i,\b_4}] \\
\chi_{2k_i}&=&[{a^\dagger}_{-k_i,\a_1}..{a^\dagger}_{-k_i,\a_4}{a^\dagger}_{-k_i,\b_1}..{a^\dagger}_{-k_i,\b_4}]
\end{eqnarray}
where, i=1,2.
Before giving the expression of $A, B$ matrix, we write the independent element which is found in $A, B$.
\begin{eqnarray}
a(k) &=&J_1\frac{1}{2}[(1+\cos 2\phi) \cos k_y-2\ cos 2\phi] \\ \nonumber
      && -(|J_2|+2|J_3|) \cos\phi,\\
b(k) &=&\frac{1}{2}J_1(1-\cos 2\phi)\cos k_y,\\
d_2(k)&=&\frac{1}{2}J_2(1+\cos\phi)\cos\frac{k_y}{2}e^{\frac{ik_y}{2}},\\
d_3(k) &=&\frac{1}{2}J_3(1+\cos\phi)\cos\frac{k_y}{2}e^{\frac{ik_y}{2}},\\
e_2(k)&=&\frac{1}{2}J_2(1-\cos\phi)\cos\frac{k_y}{2}e^{\frac{ik_y}{2}},\\
e_3(k)&=&\frac{1}{2}J_3(1-\cos\phi)\cos\frac{k_y}{2}e^{\frac{ik_y}{2}},\\
d_m(k)&=&\frac{1}{2}J_3(1+\cos\phi)\cos\frac{k_y}{2}e^{i(\frac{k_y}{2}-k_x)},\\
d_p(k)&=&\frac{1}{2}J_3(1+\cos\phi)\cos\frac{k_y}{2}e^{i(\frac{k_y}{2}+k_x)},\\
e_m(k)&=&\frac{1}{2}J_3(1-\cos\phi)\cos\frac{k_y}{2}e^{i(\frac{k_y}{2}-k_x)},\\
e_p(k)&=&\frac{1}{2}J_3(1-\cos\phi)\cos\frac{k_y}{2}e^{i(\frac{k_y}{2}+k_x)},\\
f_m(k)&=&\frac{1}{2}J_3(1+\cos\phi)\cos\frac{k_y}{2}e^{i(\frac{k_y}{2}-k_z)},\\
f_p(k)&=&\frac{1}{2}J_3(1+\cos\phi)\cos\frac{k_y}{2}e^{i(\frac{k_y}{2}+k_z)},\\
g_m(k)&=&\frac{1}{2}J_3(1-\cos\phi)\cos\frac{k_y}{2}e^{i(\frac{k_y}{2}-k_z)},\\
g_p(k)&=&\frac{1}{2}J_3(1-\cos\phi)\cos\frac{k_y}{2}e^{i(\frac{k_y}{2}+k_z)}\\
h_1&=&h/2,~~h_2= he^{i\phi}/2 .
\end{eqnarray}

\subsection{For CH phase}

\begin{eqnarray}
A(k)&=&
\begin{bmatrix}
a & d_2 & 0 & d_m & 0 & 0 & 0 & d_3 \\
{d_2}^* & a & {d_3}^* & 0 & 0 & 0 & {g_p}^* & 0 \\
0 & d_3 & a & d_2 & 0 & g_m & 0 & 0 \\
{d_m}^* & 0 & {d_2}^* & a & {d_3}^* & 0 & 0 & 0 \\
0 & 0 & 0 & d_3 & a & d_2 & 0 & d_p \\
0 & 0 & {g_m}^* & 0 & {d_2}^* & a & {d_3}^* & 0 \\
0 & {g_p} & 0 & 0 & 0 & d_3 & a & d_2 \\
{d_3}^* & 0 & 0 & 0 & {d_p}^* & 0 & {d_2}^* & a 
\end{bmatrix}
\end{eqnarray}
\begin{eqnarray}
B(k)&=&
\begin{bmatrix}
b & e_2 & 0 & e_m & 0 & 0 & 0 & e_3 \\
{e_2}^* & b & {e_3}^* & 0 & 0 & 0 & {f_p}^* & 0 \\
0 & e_3 & b & e_2 & 0 & f_m & 0 & 0 \\
{e_m}^* & 0 & {e_2}^* & b & {e_3}^* & 0 & 0 & 0 \\
0 & 0 & 0 & e_3 & b & e_2 & 0 & e_p \\
0 & 0 & {f_m}^* & 0 & {e_2}^* & b & {e_3}^* & 0 \\
0 & {f_p} & 0 & 0 & 0 & e_3 & b & e_2 \\
{e_3}^* & 0 & 0 & 0 & {e_p}^* & 0 & {e_2}^* & b 
\end{bmatrix}
\end{eqnarray}
\begin{eqnarray}
F&=&
\begin{bmatrix}
h_1 & 0 & 0 & 0 & 0 & 0 & 0 & 0 \\
0 & h_2 & 0 & 0 & 0 & 0 & 0 & 0 \\
0 & 0 & h_1 & 0 & 0 & 0 & 0 & 0 \\
0 & 0 & 0 & h_2 & 0 & 0 & 0 & 0 \\
0 & 0 & 0 & 0 & h_1 & 0 & 0 & 0 \\
0 & 0 & 0 & 0 & 0 & h_2 & 0 & 0 \\
0 & 0 & 0 & 0 & 0 & 0 & h_1 & 0 \\
0 & 0 & 0 & 0 & 0 & 0 & 0 & h_2\\
\end{bmatrix}
\end{eqnarray}
\subsection{For C2H phase}
\begin{eqnarray}
A(k)&=&
\begin{bmatrix}
a & d_3 & 0 & 0 & 0 & d_m & 0 & e_2 \\
{d_3}^* & a & {e_2}^* & 0 & {d_p}^* & 0 & 0 & 0 \\
0 & e_2 & a & d_3 & 0 & 0 & 0 & g_m \\
0 & 0 & {d_3}^* & a & {e_2}^* & 0 & {g_p}^* & 0 \\
0 & d_p & 0 & e_2 & a & d_3 & 0 & 0 \\
{d_m}^* & 0 & 0 & 0 & {d_3}^* & a & {e_2}^* & 0 \\
0 & 0 & 0 & g_p & 0 & e_2 & a & d_3 \\
{e_2}^* & 0 & {g_m}^* & 0 & 0 & 0 & {d_3}^* & a 
\end{bmatrix}
\end{eqnarray}
\begin{eqnarray}
B(k)&=&
\begin{bmatrix}
b & e_3 & 0 & 0 & 0 & e_m & 0 & d_2 \\
{e_3}^* & b & {d_2}^* & 0 & {e_p}^* & 0 & 0 & 0 \\
0 & d_2 & b & e_3 & 0 & 0 & 0 & f_m \\
0 & 0 & {e_3}^* & b & {d_2}^* & 0 & {f_p}^* & 0 \\
0 & e_p & 0 & d_2 & b & e_3 & 0 & 0 \\
{e_m}^* & 0 & 0 & 0 & {e_3}^* & b & {d_2}^* & 0 \\
0 & 0 & 0 & f_p & 0 & d_2 & b & e_3 \\
{d_2}^* & 0 & {f_m}^* & 0 & 0 & 0 & {e_3}^* & b 
\end{bmatrix}
\end{eqnarray}
\begin{eqnarray}
F&=&
\begin{bmatrix}
h_1 & 0 & 0 & 0 & 0 & 0 & 0 & 0 \\
0 & h_2 & 0 & 0 & 0 & 0 & 0 & 0 \\
0 & 0 & -h_1 & 0 & 0 & 0 & 0 & 0 \\
0 & 0 & 0 & -h_2 & 0 & 0 & 0 & 0 \\
0 & 0 & 0 & 0 & h_1 & 0 & 0 & 0 \\
0 & 0 & 0 & 0 & 0 & h_2 & 0 & 0 \\
0 & 0 & 0 & 0 & 0 & 0 & -h_1 & 0 \\
0 & 0 & 0 & 0 & 0 & 0 & 0 & -h_2\\
\end{bmatrix}
\end{eqnarray}
\subsection{A2H phase}
\begin{eqnarray}
A(k)&=&
\begin{bmatrix}
a & e_3 & 0 & 0 & 0 & e_m & 0 & d_2 \\
{e_3}^* & a & {d_2}^* & 0 & {e_p}^* & 0 & 0 & 0 \\
0 & d_2 & a & e_3 & 0 & 0 & 0 & f_m \\
0 & 0 & {e_3}^* & a & {d_2}^* & 0 & {f_p}^* & 0 \\
0 & e_p & 0 & d_2 & a & e_3 & 0 & 0 \\
{e_m}^* & 0 & 0 & 0 & {e_3}^* & a & {d_2}^* & 0 \\
0 & 0 & 0 & f_p & 0 & d_2 & a & e_3 \\
{d_2}^* & 0 & {f_m}^* & 0 & 0 & 0 & {e_3}^* & a 
\end{bmatrix}
\end{eqnarray}

\begin{eqnarray}
B(k)&=&
\begin{bmatrix}
b & d_3 & 0 & 0 & 0 & d_m & 0 & e_2 \\
{d_3}^* & b & {e_2}^* & 0 & {d_p}^* & 0 & 0 & 0 \\
0 & e_2 & b & d_3 & 0 & 0 & 0 & g_m \\
0 & 0 & {d_3}^* & b & {e_2}^* & 0 & {g_p}^* & 0 \\
0 & d_p & 0 & e_2 & b & d_3 & 0 & 0 \\
{d_m}^* & 0 & 0 & 0 & {d_3}^* & b & {e_2}^* & 0 \\
0 & 0 & 0 & g_p & 0 & e_2 & b & d_3 \\
{e_2}^* & 0 & {g_m}^* & 0 & 0 & 0 & {d_3}^* & b 
\end{bmatrix}
\end{eqnarray}
\begin{eqnarray}
F&=&
\begin{bmatrix}
h_1 & 0 & 0 & 0 & 0 & 0 & 0 & 0 \\
0 & -h_2 & 0 & 0 & 0 & 0 & 0 & 0 \\
0 & 0 & -h_1 & 0 & 0 & 0 & 0 & 0 \\
0 & 0 & 0 & h_2 & 0 & 0 & 0 & 0 \\
0 & 0 & 0 & 0 & h_1 & 0 & 0 & 0 \\
0 & 0 & 0 & 0 & 0 & -h_2 & 0 & 0 \\
0 & 0 & 0 & 0 & 0 & 0 & -h_1 & 0 \\
0 & 0 & 0 & 0 & 0 & 0 & 0 & h_2\\
\end{bmatrix}
\end{eqnarray}
\subsection{FH phase}

\begin{eqnarray}
A(k)&=&
\begin{bmatrix}
a & e_3 & 0 & 0 & 0 & e_m & 0 & e_2 \\
{e_3}^* & a & {e_2}^* & 0 & {e_p}^* & 0 & 0 & 0 \\
0 & e_2 & a & e_3 & 0 & 0 & 0 & f_m \\
0 & 0 & {e_3}^* & a & {e_2}^* & 0 & {f_p}^* & 0 \\
0 & e_p & 0 & e_2 & a & e_3 & 0 & 0 \\
{e_m}^* & 0 & 0 & 0 & {e_3}^* & a & {e_2}^* & 0 \\
0 & 0 & 0 & f_p & 0 & e_2 & a & e_3 \\
{e_2}^* & 0 & {f_m}^* & 0 & 0 & 0 & {e_3}^* & a 
\end{bmatrix}
\end{eqnarray}
\begin{eqnarray}
B(k)&=&
\begin{bmatrix}
b & d_3 & 0 & 0 & 0 & d_m & 0 & d_2 \\
{d_3}^* & b & {d_2}^* & 0 & {d_p}^* & 0 & 0 & 0 \\
0 & d_2 & b & d_3 & 0 & 0 & 0 & g_m \\
0 & 0 & {d_3}^* & b & {d_2}^* & 0 & {g_p}^* & 0 \\
0 & d_p & 0 & d_2 & b & d_3 & 0 & 0 \\
{d_m}^* & 0 & 0 & 0 & {d_3}^* & b & {d_2}^* & 0 \\
0 & 0 & 0 & g_p & 0 & d_2 & b & d_3 \\
{d_2}^* & 0 & {g_m}^* & 0 & 0 & 0 & {d_3}^* & b 
\end{bmatrix}
\end{eqnarray}
\begin{eqnarray}
F=
\begin{bmatrix}
h_1 & 0 & 0 & 0 & 0 & 0 & 0 & 0 \\
0 & -h_2 & 0 & 0 & 0 & 0 & 0 & 0 \\
0 & 0 & h_1 & 0 & 0 & 0 & 0 & 0 \\
0 & 0 & 0 & -h_2 & 0 & 0 & 0 & 0 \\
0 & 0 & 0 & 0 & h_1 & 0 & 0 & 0 \\
0 & 0 & 0 & 0 & 0 & -h_2 & 0 & 0 \\
0 & 0 & 0 & 0 & 0 & 0 & h_1 & 0 \\
0 & 0 & 0 & 0 & 0 & 0 & 0 & -h_2\\
\end{bmatrix}
\end{eqnarray}

\begin{widetext}
\section{Appendix-B, Symmetry Point Eigenvalues}

\subsubsection{CH phase}

\begin{itemize}
\item $\G$-point energies:

$0,\sqrt{(a\pm J_2d)^2- (b\pm J_2e)^2 },$ \\
$\sqrt{(a\pm (J_2-2J_3)d)^2- (b\pm (J_2-2J_3)e)^2},$ \\
$\sqrt{(a- (J_2+2J_3)d)^2- (b- (J_2+2J_3)e)^2}$

\item $X$-point energies:

$\sqrt{a^2-b^2+(d^2-e^2)({J_2}^2+2{J_3}^2)-2J_3(ad-be)\pm  2\sqrt{(J_3(d^2-e^2)-(ad-be))^2({J_2}^2+{J_3}^2)} }, \\ \sqrt{a^2-b^2+(d^2-e^2)({J_2}^2+2{J_3}^2)-2J_3(ad-be)\pm   2\sqrt{(J_3(d^2-e^2)+(ad-be))^2({J_2}^2+{J_3}^2)}} $

\item $R$-point energies:

$\sqrt{(a\pm J_2d)^2- (b\pm J_2e)^2 }, \sqrt{a^2-b^2+(d^2-e^2)({J_2}^2+4{J_3}^2) \pm 2\sqrt{(ad-be)^2({J_2}^2+4{J_3}^2)}}$

\end{itemize}

\subsubsection{C2H phase}

\begin{itemize}
\item $\G$-point energies:

$0,\sqrt{(a\pm J_2e)^2- (b\pm J_2d)^2 }, \sqrt{(a-J_2e)^2-(b-J_2d)^2-4J_3(ad-be-J_3(d^2-e^2))},\\ \sqrt{(a\pm J_2e)^2-(b\pm J_2d)^2\mp 4J_3(ad-be\mp J_3(d^2-e^2))}$

\item $X$-point energies:

$\sqrt{  \begin{aligned} a^2-b^2-(d^2-e^2)({J_2}^2-2{J_3}^2)-2J_3(ad-be)\pm \\ 2\sqrt{(d^2-e^2)(2{J_3}^2(be-ad)+(d^2-e^2){J_3}^2(-{J_2}^2+{J_3}^2))+{J_3^2}(ad-be)^2+{J_2^2}(ae-bd)^2}  \end{aligned}}, \\ \sqrt{  \begin{aligned} a^2-b^2-(d^2-e^2)({J_2}^2-2{J_3}^2)+2J_3(ad-be)\pm \\ 2\sqrt{(d^2-e^2)(2{J_3}^2(ad-be)+(d^2-e^2){J_3}^2(-{J_2}^2+{J_3}^2))+{J_3^2}(ad-be)^2+{J_2^2}(ae-bd)^2}  \end{aligned}} $

\item $R$-point energies:

$\sqrt{(a\pm J_2e)^2- (b\pm J_2d)^2 },$ \\
$ \sqrt{a^2-b^2+(d^2-e^2)(-{J_2}^2+4{J_3}^2) \pm 2\sqrt{(ae-bd)^2{J_2}^2+4{J_3}^2(ad-be)^2-4{J_2}^2{J_3}^2(d^2-e^2)}}$

\end{itemize}
\subsubsection{A2H phase}

\begin{itemize}
\item $\G$-point energies:

$0,\sqrt{(a\pm J_2d)^2- (b\pm J_2e)^2 }, \sqrt{(a-J_2d)^2-(b-J_2e)^2-4J_3(ae-bd+J_3(d^2-e^2))},\\ \sqrt{(a\pm J_2d)^2-(b\pm J_2e)^2\mp 4J_3(ae-bd\mp J_3(d^2-e^2))}$

\item $X$-point energies:

$\sqrt{  \begin{aligned} a^2-b^2+(d^2-e^2)({J_2}^2-2{J_3}^2)-2J_3(ae-bd)\pm \\ 2\sqrt{(d^2-e^2)(2{J_3}^3(ae-bd)+(d^2-e^2){J_3}^2(-{J_2}^2+{J_3}^2))+{J_2^2}(ad-be)^2+{J_3^2}(ae-bd)^2}  \end{aligned}}, \\ \sqrt{  \begin{aligned} a^2-b^2-(d^2-e^2)({J_2}^2-2{J_3}^2)+2J_3(ae-bd)\pm \\ 2\sqrt{(d^2-e^2)(2{J_3}^3(bd-ae)+(d^2-e^2){J_3}^2(-{J_2}^2+{J_3}^2))+{J_2^2}(ad-be)^2+{J_3^2}(ae-bd)^2}  \end{aligned}} $

\item $R$-point energies:

$\sqrt{(a\pm J_2e)^2- (b\pm J_2d)^2 }, $ \\
$\sqrt{a^2-b^2+(d^2-e^2)({J_2}^2-4{J_3}^2) \pm 2\sqrt{(ad-be)^2{J_2}^2+4{J_3}^2(ae-bd)^2-4{J_2}^2{J_3}^2(d^2-e^2)}}$

\end{itemize}

\subsubsection{FH phase}
\begin{itemize}
\item $\G$-point energies:

$0,\sqrt{(a\pm J_2e)^2- (b\pm J_2d)^2 }, $ \\
$\sqrt{(a\pm (J_2-2J_3)e)^2- (b\pm (J_2-2J_3)d)^2}, \sqrt{(a- (J_2+2J_3)e)^2- (b- (J_2+2J_3)d)^2}$

\item $X$-point energies:

$\sqrt{a^2-b^2-(d^2-e^2)({J_2}^2+2{J_3}^2)-2J_3(ae-bd)\pm  2\sqrt{(J_3(d^2-e^2)+(ae-bd))^2({J_2}^2+{J_3}^2)} }, \\ \sqrt{a^2-b^2-(d^2-e^2)({J_2}^2+2{J_3}^2)+2J_3(ae-bd)\pm  2\sqrt{(J_3(d^2-e^2)-(ae-bd))^2({J_2}^2+{J_3}^2)} } $

\item $R$-point energies:

$\sqrt{(a\pm J_2d)^2- (b\pm J_2e)^2 }, \sqrt{a^2-b^2+(d^2-e^2)({J_2}^2+4{J_3}^2) \pm 2\sqrt{(ad-be)^2({J_2}^2+4{J_3}^2)}}$

\end{itemize}

In each phase Z-point has same energies as X-point has. In each case M-point has energy $\sqrt{(J_1+E_{GS})(J_1\cos 2\phi +E_{GS})}$. where $a=\frac{J_1}{2}(1-\cos 2\phi)-(|J_2|+2|J_3|)\cos\phi$, $b=\frac{1}{2}J_1(1-\cos 2\phi)$, $d=\frac{1}{2}(1+\cos\phi)$, $e=\frac{1}{2}(1-\cos\phi)$,  and \\$E_{GS}=J_1\cos 2\phi +(|J_2|+|J_3|)\cos\phi$
\end{widetext}

\section{Appendix-C, Derivation of Susceptibilities}

For a single ladder and with field direction lying on the plan of polarization (say, $x$-direction):\\
For $\alpha$-chain the perturbing Hamiltonian is-
\begin{eqnarray}
{H^\prime}_{\alpha} &=&-\sum_{m}h{S^x}_{\alpha} \nonumber \\
&=&-h\sum_{m}\left(\cos{(2m\phi)}{S^{x^\prime}}_{\alpha}-\sin(2m\phi){S^{z^\prime}}_{\alpha}\right) \nonumber \\
&=& \frac{h}{2}\sum_{k}\left({a^\dagger}_{k,\alpha_j}  {a}_{k+2\phi,\alpha}+{a^\dagger}_{k+2\phi,\alpha}  {a}_{k,\alpha}\right)\nonumber \\
&-&\frac{h}{2}\sqrt{\frac{s}{2}}\left({a^\dagger}_{2\phi , \alpha}-{a^\dagger}_{-2\phi , \alpha}+a_{2\phi , \alpha}-a_{-2\phi , \alpha}\right)
\end{eqnarray}
Similarly for $\beta$-chain this Hamiltonian can be derived by  replacing $2m\phi$ by $(2m+1)\phi$, i.e., ${H^\prime}_{\beta}=\left({H^\prime}_{\alpha}\right)_{2m\phi\rightarrow(2m+1)\phi}$.

Therefore,
\begin{eqnarray}
{H^\prime}_{\beta}&=&-\sum_{m}h{S^x}_{\beta} \nonumber \\
&=&-h\sum_{m}\left(\cos{(2m+1)\phi}{S^{x^\prime}}_{\beta}-\sin{(2m+1)\phi}{S^{z^\prime}}_{\beta}\right) \nonumber \\
&=&\frac{h}{2}\sum_{k}\left(e^{i\phi}{a^\dagger}_{k,\beta}  {a}_{k+2\phi,\beta}+e^{-i\phi}{a^\dagger}_{k+2\phi,\beta}  {a}_{k,\beta}\right) \nonumber \\
&-&\frac{h}{2}\sqrt{\frac{s}{2}}\left(e^{-i\phi}{a^\dagger}_{2\phi , \beta}-e^{i\phi}{a^\dagger}_{-2\phi , \beta}\right)
\nonumber \\
&-&\frac{h}{2}\sqrt{\frac{s}{2}}\left(e^{i\phi}a_{2\phi , \beta}-e^{-i\phi}a_{-2\phi , \beta}\right)
\end{eqnarray}
The quadratic terms in the perturbing  Hamiltonian can be absorbed in the free Hamiltonian  and that can be diagonalized using a transformation of the basis[Eq.~\ref{hpara}] (with a displacement in $k$ as $k\rightarrow k-\phi$ to maintain the same positive and negative eigenvalues).
With the new basis we can write the remaining perturbing terms as:
\begin{multline}
{H^\prime}/ \left(\frac{h}{2}\sqrt{\frac{s}{2}}\right)={a^\dagger}_{2\phi , \alpha}-{a^\dagger}_{-2\phi , \alpha}+a_{2\phi , \alpha}-a_{-2\phi , \alpha}\\+e^{-i\phi}{a^\dagger}_{2\phi , \beta}-e^{i\phi}{a^\dagger}_{-2\phi , \beta}+e^{i\phi}a_{2\phi , \beta}-e^{-i\phi}a_{-2\phi , \beta}\\
=\left(\sum_{j=1,4}{b^\dagger}_{\phi,j} +\sum_{j=5,8}b_{-\phi,j}\right)q_j
\end{multline}

Where, $q_j=({-t^\star}_{j,3}+t_{7,j}-e^{i\phi}{t^\star}_{j,4}+e^{i\phi}t_{8,j}+h.c.)_{k=\phi}$. Now the Hull Hamiltonian in diagonalized basis can be written as: $H=H_0+H^\prime$. Where, $H_0=\sum_{k,j}\epsilon_{k,j}{b^\dagger}_{k,j}  b_{k,j}$. Now to remove the linear terms let us make the following transformation:
$b_{k,j}\rightarrow b_{k,j}+p_j$. Now a very straight forward calculation shows that  the linear terms will be completely removed for $p_j=-\frac{h}{2}\sqrt{\frac{s}{2}} q_j\delta_{k,\phi}/\epsilon_{\phi,j}$. Thus the energy correction for the linear terms can be written as:
\begin{equation}
E^\prime=-\sum_j \frac{1}{\epsilon_{\phi,j}} p_j{p_j}^\star\delta_{k,0}
\end{equation} 

Now let us calculate the magnetization:

$M_{\|}=\left<S^x\right>=M_1+M_2$ where,

$M_1=$
$-< \frac{1}{2}\sum_k({a^\dagger}_{k-\phi,\alpha_j}  {a}_{k+\phi,\alpha}+e^{i\phi}{a^\dagger}_{k-\phi,\beta}  {a}_{k+\phi,\beta}\\+h.c.)>=-\frac{1}{2}\sum_{k,j}<{b^\dagger}_{k,j}  b_{k,j}>C_j$. Therefore,

\begin{equation}
\Rightarrow M_1=-\frac{1}{2}\sum_{k,j}<n_{k,j}>C_j
\end{equation}
Where, $C_j=t_{1,j}{t^\star}_{j,3}+e^{-i\phi}t_{2,j}{t^\star}_{j,4}+h.c.$ 
and $t$-elements form the unitary transformation matrix that transforms $'a'-basis$ to $diagonalized$ $'b'-basis$ $\&$ the bosonic distribution function,
\begin{equation}
 n_{k,j}=\frac{1}{e^{\beta \epsilon_{k,j}}-1}
\end{equation}

And,


$M_2=\frac{1}{2}\sqrt{\frac{s}{2}} <({a^\dagger}_{2\phi , \alpha}-{a^\dagger}_{-2\phi , \alpha}+e^{-i\phi}{a^\dagger}_{2\phi , \beta}-e^{i\phi}{a^\dagger}_{-2\phi , \beta}+h.c.)>$
$=-<\frac{H^\prime}{h}>$. Therefore,
\begin{equation}
M_2=\sqrt{\frac{s}{2}} \sum_{j=1,8}\frac{1}{\epsilon_{\phi,j}}p_j{p_j}^\star\delta_{k,\phi}
\end{equation}

\vspace{0.25cm}
All these calculations have been done for a single $J_1-J_2$ (Fig 1) we extended this calculation for full Hollandite case and found out the magnetic susceptibility curves.

\end{document}